\documentclass[superscriptaddress]{revtex4-1}

\usepackage{graphicx,psfrag}
\usepackage{epstopdf}
\usepackage{amsmath}
\usepackage{bm}
\usepackage{amssymb}
\usepackage{subfigure}

\begin{document}
\title[]{Design of an underwater acoustic bend by pentamode metafluid}

\author{Zhaoyong Sun}
\email{sunzhaoyong11@mails.ucas.ac.cn}
\affiliation{Key Laboratory of Noise and Vibration Research, Institute of Acoustics,
Chinese Academy of Sciences, Beijing 100190, People's Republic of China}
\affiliation{University of Chinese Academy of Sciences, Beijing 100049, People’s Republic of China}
\author{Han Jia}
\email{hjia@mail.ioa.ac.cn}
\affiliation{Key Laboratory of Noise and Vibration Research, Institute of Acoustics,
Chinese Academy of Sciences, Beijing 100190, People's Republic of China}
\affiliation{University of Chinese Academy of Sciences, Beijing 100049, People’s Republic of China}
\affiliation{State Key Laboratory of Acoustics, Institute of Acoustics, Chinese Academy of Sciences, Beijing 100190,
People’s Republic of China}
\author{Yi Chen}
\email{chenyi221@gmail.com}
\affiliation{School of Aerospace Engineering, Beijing Institute of Technology, Beijing 100081, People's Republic of China}
\author{Zhen Wang}
\email{wangzhen1@mails.ioa.ac.cn}
\affiliation{Key Laboratory of Noise and Vibration Research, Institute of Acoustics,
Chinese Academy of Sciences, Beijing 100190, People's Republic of China}
\affiliation{University of Chinese Academy of Sciences, Beijing 100049, People’s Republic of China}

\author{Jun Yang}
\email{jyang@mail.ioa.ac.cn}
\affiliation{Key Laboratory of Noise and Vibration Research, Institute of Acoustics,
Chinese Academy of Sciences, Beijing 100190, People's Republic of China}
\affiliation{University of Chinese Academy of Sciences, Beijing 100049, People’s Republic of China}
\affiliation{State Key Laboratory of Acoustics, Institute of Acoustics, Chinese Academy of Sciences, Beijing 100190,
People’s Republic of China}


\begin{abstract}    
We design an impedance matching underwater acoustic bend with pentamode microstructure. 
The proposed bend is assembled by pentamode lattice. 
The effective density and compressive modulus of each unit cell can be tuned simultaneously, which are modulated to guarantee both the bending effect and high transmission.
The standard deviations of transmitted phase are calculated to quantitatively evaluate the degree of the distortion of the transmitted wavefront, while the transmission is calculated to appraise the degree of acoustic impedance matching.
The low standard deviations and high transmission indicate that the designed bend has a nice broadband bending effect and is impedance-matched to water.
This design has potential applications in underwater communication and underwater detection.
\end{abstract}

\maketitle


\section{\label{intro}Introduction}
Acoustic metamaterials can exhibit many abnormal acoustic properties, such as negative density, negative modulus\cite{liu2000locally,mei2007effective,lee2010composite}, anisotropic density\cite{torrent2010anisotropic,christensen2012anisotropic} and anisotropic modulus\cite{shen2014anisotropic,kadic2013anisotropic}. 
These remarkable properties provide possibilities to manipulate sonic waves in ways that are impossible in ordinary materials.
With these abnormal properties, acoustic metamaterials can be used in acoustic cloaking\cite{chen2017broadband,hu2013an,bi2017design,pendry2008acoustic}, acoustic subwavelength imaging\cite{kaina2015negative,park2015acoustic,jia2010subwavelength,deng2009theoretical}, rainbow trapping\cite{ni2014acoustic,zhou2016precise,jia2014spatial} and so on.

Bending the wave without distorting its wavefront is one of these interesting applications. 
In recent years, some interesting work about acoustic bend has been reported. 
L.Wu $et al.$ employed a two-dimensional graded sonic crystal to realize a graded index medium based acoustic bending waveguide\cite{wu2011acoustic}. 
Y. Wang $et al.$ proposed an anisotropic metamaterial with only one component of the mass density tensor near zero to control the sound wave propagation and  realized perfect bending waveguides numerically\cite{wang2016acoustic}.
W. Lu et al. have designed and fabricated a broadband acoustic right angle bend by perforated panels which can tune the effective density\cite{lu_design_2017}. 
In our previous work\cite{yang2017impedance}, we used perforated plates with side pipes to design an acoustic bend working in air. 
The effective bulk modulus is tuned by the pipes, while the effective density is adjusted by the perforated panels simultaneously. 
This enables the bend to match acoustic impedance to air. 
Acoustic bend also have important applications in underwater communication and underwater detection.
However, perforated panels and the side pipes can not be used in water, since it is difficult to find a rigid material in water. 
Therefore, it is necessary to consider other approaches. 
An appropriate option is  pentamode material(PM) \cite{milton1995elasticity}. 
PM is a kind of special solid structure with tunable effective modulus and density that  only guarantees the longitudinal wave to propagate in it, and can be designed to have anisotropic modulus\cite{kadic2013anisotropic,layman2013highly}. 
This makes it have the similar acoustic property with fluid\cite{norris2009acoustic} and the advantage for designing impedance matching underwater devices\cite{su2017broadband,tian2015broadband}.

In this article, we use 2-dimensional(2D) version of the pentamode material(PM) lattice to design an underwater  acoustic bend.
The required acoustic parameters are obtained by  theoretical calculation.
From the calculation, the modulus of the bend is proportional to the radial position, while the density is inversely proportional to the radial position. 
This property is the key point to protect the wavefront from being distorted and ensure the acoustic impedance matched  to water. 
The unit cells are designed by a homogenization method\cite{chen_latticed_2015}, and their effective parameters  coincide with the required parameters well.
The latticed pentamode bend(PMB) is  assembled by these unit cells. 
The simulated results are demonstrated and show the  broadband bending effect and high transmission of the latticed PMB. 
\section{\label{bend}Theory of the acoustic bend}
\begin{figure}[!ht]
\begin{center}
\includegraphics[width =2.5in]{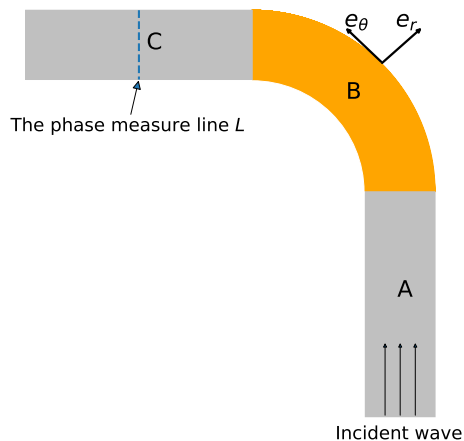}
\caption{\label{fig:FIG1}(Color online) A schematic view of the acoustic bend with waveguides. Domain B is the acoustic bend, while domain A and C are a horizontal and a vertical waveguide correspondingly.}
\end{center}	
\end{figure}

Figure \ref{fig:FIG1} illustrates the outline of the 2D   acoustic bend. 
In the proposed model, an acoustic bend (domain B) connects a vertical waveguide (domain A) with a horizontal waveguide (domain C).
The inner and outer radii are denoted as $r_1$ and $r_2$. 
These two waveguides are filled with the background fluid with density $\rho_0$ and bulk modulus $K_0$, which give the sound velocity $c_0=\sqrt{\frac{K_0}{\rho_0}}$.
The wave is emitted from the bottom of the vertical waveguide, and assumed to travel with the velocity $c_\theta=c_\theta (r)$ along the  azimuthal direction in the bend.

For an ideal acoustic bend, two conditions have to be satisfied in order to keep the nice bending effect and high transmission: one is that the phase of the wave does not vary along the radial direction, and the other is that all energy of the incident wave can pass through the bend. 
The former implies that the travel time of the wave along different curves $r=r_i$ is the same, while the latter signifies that the acoustic impedance of the bend matches to the fluid in the waveguides. 
These two conditions can immediately give the following equations: 
\begin{equation}
\frac{r_i\theta}{c_{\theta}(r_i)}=\frac{r_j\theta}{c_{\theta}(r_j)},~ ~ \rho_B(r) c_\theta(r)=\rho_0 c_0
\label{eq:bend}
\end{equation}
where $r_i$ and $r_j$ are two arbitrary radial coordinates, $\rho_B(r)$ is the density of the bend. 
Equation (\ref{eq:bend}) shows that velocity $c_\theta(r)$  is proportional to the radial coordinate, while the density $\rho_B(r)$ is inversely proportional to the radial coordinate. 
This gives the acoustic parameters distributions of the bend as follows: 
\begin{equation}    
    \rho_B(r)=\frac{b\rho_0}{r},~  K=\frac{rK_0}{b}
    \label{eq:bendpara}
\end{equation}
where $K$ is compressive modulus of the bend B($K=\rho_B c_\theta^2$), and  $b$ is a constant that can tune the required density and modulus of the bend.  

Since Eq.(\ref{eq:bend}) and Eq.(\ref{eq:bendpara}) are derived by the analysis in azimuthal direction, the acoustic parameters along the radial direction does not influence the bending effect. 
In the work of the acoustic bend in air \cite{yang2017impedance}, we used the unit cells with anisotropic density and isotropic modulus to design the bend structure. 
The azimuthal density and the bulk modulus satisfy Eq.(\ref{eq:bendpara}), which makes the impedance match to air. 

Here, we use 2D PMs with anisotropic modulus and isotropic density to design an underwater PM bend(PMB). 
The acoustic parameters of the PMB are normalized to the corresponding values of water, and shown as follows 
\begin{equation}    
\rho=\frac{b}{r},~  K_\theta=\frac{r}{b},~  K_r=\frac{K_\theta}{2},~b=0.6 m
    \label{eq:property}
\end{equation}
with the normalized azimuthal compressive modulus $K_\theta$, normalized radial compressive modulus $K_r$ and the normalized density $\rho$.
\begin{figure*}[!htbp] 
\includegraphics[width=0.8\textwidth]{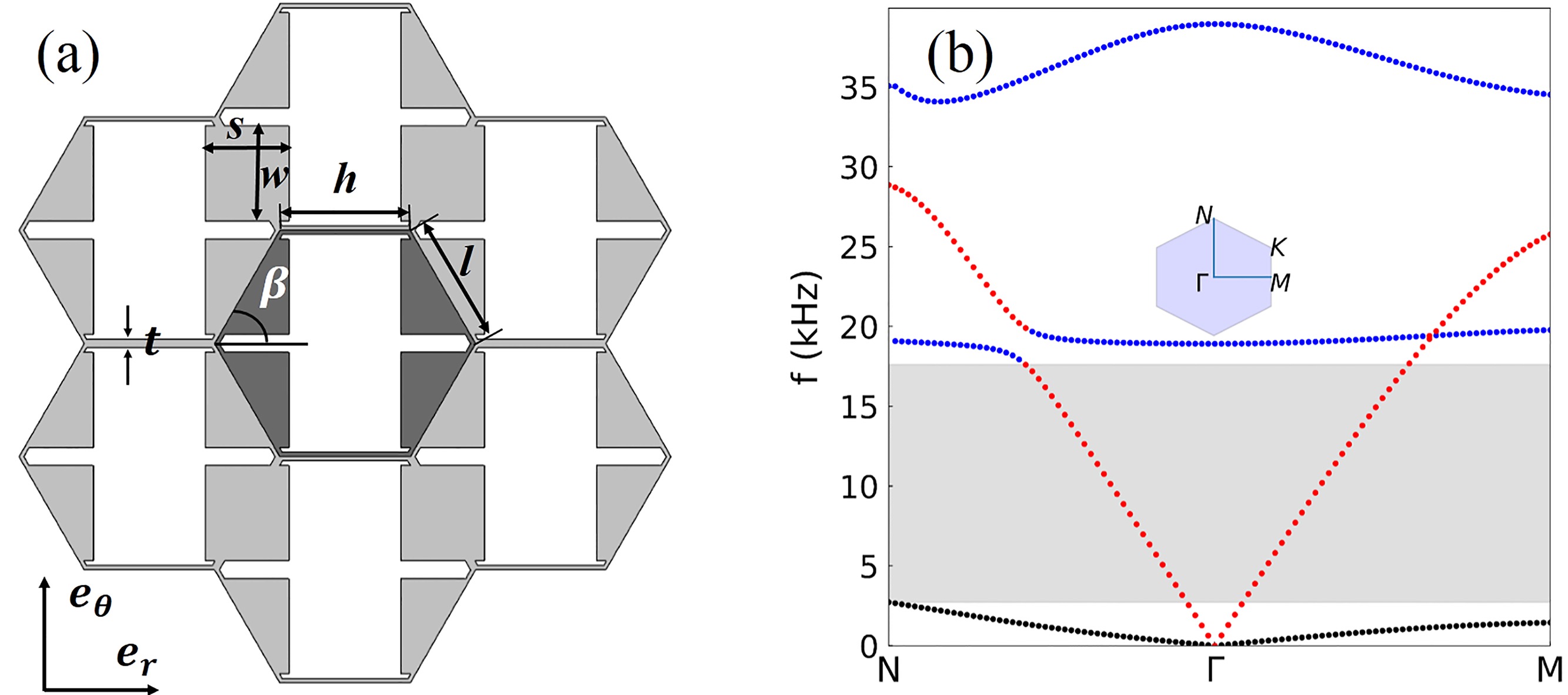}
\caption{\label{fig:FIG2}(Color online) (a) Microstructure of 2D PMs, with a highlighted unit cell. (b) Band diagram along the $N-\Gamma-M$ direction. The first Brillouin zone is shown  in the center of the figure.}
\end{figure*}   
\section{\label{PM}Design of the unit cells for the latticed pentamode material bend}
The elastic tensor of the 2D PM that used in the underwater bend can be characterized as follows in the polar coordinates: 
 \begin{equation}
 \mathbf{C} = 
 \left(\begin{array}{ccc}
 C_{11}&C_{12}&0\\
 C_{12}&C_{22}&0\\
 0&0&C_{33}\\
 \end{array} \right)
=K_0\left(\begin{array}{ccc}
 K_{r}&K_{r\theta}&0\\
 K_{r\theta}&K_{\theta}&0\\
 0&0&G_{r\theta}\\
 \end{array} \right)
\label{eq:elastensor}
\end{equation}
where $K_0$ is the bulk modulus of the background fluid. 
The background fluid is water with density $\rho_0=1000$ kg/m$^3$ and bulk modulus $K_0=2.25$ GPa. 
For a perfect PM, the matrix elements in Eq.(\ref{eq:elastensor}) have to satisfy the condition $K_rK_\theta=K_{r\theta}^2$ and $G_{r\theta}=0$ so that the modes pertaining shear deformations vanish and only the compressive one is reserved. 
In this sense, only longitudinal wave can propagate in the perfect PM\cite{norris2009acoustic}. 
In practical case, the shear modulus   can not be zero because of the engineering limit and the demand for structural stability\cite{kadic2013anisotropic,layman2013highly}. 
Thus, the 2D practical PM needs  $K_{rr}K_{\theta\theta} \approx K_{r\theta}^2$ and $G_{r\theta} \approx 0$ to minimize the  shear modulus.

The model of 2D PM is designed as shown in Fig.\ref{fig:FIG2}.
The geometric parameters of the unit cell are characterized by length of the horizontal struts $h$, length of the oblique struts $l$, strut thickness $t$, angle $\beta$ and the block size $(\omega,s)$. 
In order to demonstrate the fluid-like property of the structure, we choose a typical unit cell and calculate the band structure using Bloch-Floquet analysis\cite{kutsenko2017wave} in COMSOL Multiphysics. 
The geometric parameters of the unit cell are set as $l=h=13.4$ mm, $\beta=60^\circ$, $t=0.7$ mm, $\omega=10$ mm and $s=0.77$ mm. 
The substrate of the latticed PM is chosen as aluminum with density $\rho_{Al}=2700$ kg/m$^3$, Young’s modulus $E_{Al}=69$ GPa and Poisson's ratio $\nu=0.33$. 
Figure \ref{fig:FIG2} depicts the first four bands. 
The black line describes the shear wave mode, while the red line represents the longitudinal wave mode.  
The two blue lines correspond to flexural modes\cite{cai2016mechanical}, which exist at high frequency range in our model. 
\begin{figure*}[ht] 
\begin{center}
\includegraphics[width=0.9\textwidth]{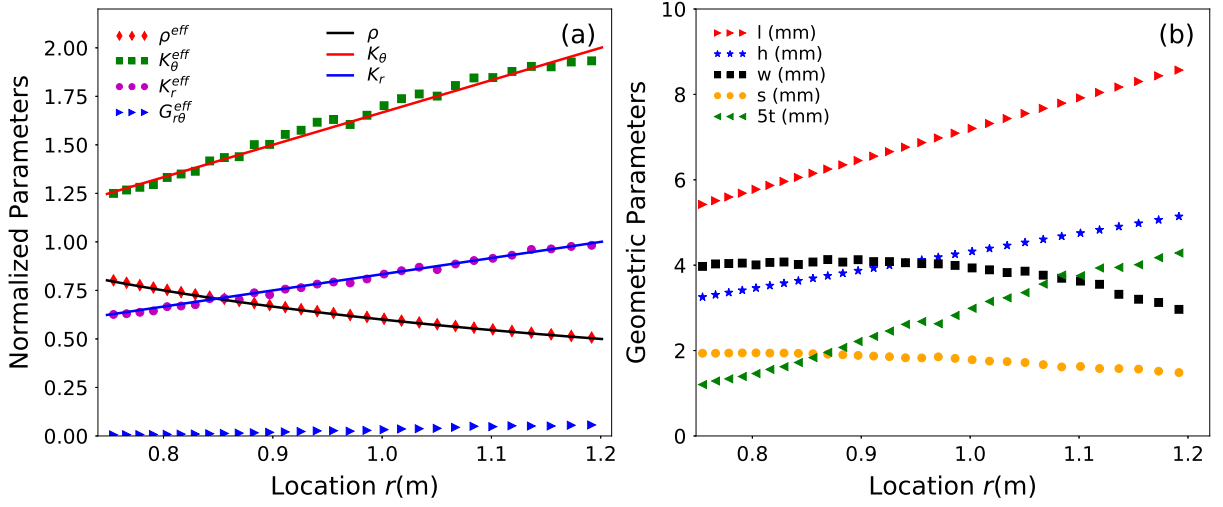}
\caption{\label{fig:FIG3}(Color online) (a) Profiles of continuously varying (solid lines) acoustic parameters  of the perfect PMB and layered (discrete symbols) objective acoustic parameters of the designed PMB. (b) The  geometric parameters of the required unit cells.}	
\end{center}
\end{figure*}

This is because the blocks on the oblique struts can prevent the struts from being bended. 
It is easy to see that at a large frequency domain (gray part), only the longitudinal modes exist in the structure, which imply the fluid-like property of the PM lattice. 
At the quasi-static regime\cite{norris2014mechanics}, the effective density of the PM lattice equals to the volume average of the unit cell mass.
Hence the effective density is immediately obtained as $\rho=1.004$ (normalized to the density of water) from the geometric parameters and the substrate material. 
The longitudinal wave velocity along the azimuthal and radial direction can be obtained from the band diagram, and read as $c_{L\theta}=1508$ m/s and $c_{Lr}=1499$ m/s. 
Thus the normalized azimuthal and radial compressive modulus  are $K_\theta=\rho\rho_0c_{L\theta}^2/K_0=1.015$  and $K_r=\rho\rho_0c_{Lr}/K_0=1.003$.
It can be seen that the effective property of this unit cell coincide with that of water well.
Therefore, for any given unit cell, the effective compressive modulus and density can be estimated with the band diagram.

The geometry of the unit cell has an important impact on the effective acoustic parameters in such a way: angle $\beta$ mainly determines the anisotropy of effective modulus along with the length ratio $\eta=\frac{h}{l}$; strut thickness $t$  primarily contributes to the effective stiffness; The block size $(\omega,s)$  can adjust the effective density almost without influencing the effective modulus.  
Thus, the required acoustic property in Eq.(\ref{eq:property}) can be realized by adjusting the geometric parameters of the unit cells and retrieving their bands.

The required acoustic parameters in Eq.(\ref{eq:property}) are depicted by solid lines in Fig.\ref{fig:FIG3}a. 
According to Eq.(\ref{eq:property}), they are symmetric in the azimuthal direction and continuous in the radial direction. 
Thus domain B in  Fig.\ref{fig:FIG1} is divided into 125 same sectors along circumferential direction.  
The continuous parameters has to be discretized with layered approximation. 
The first layer start from the location at the inner radius $r_1=743.8$mm. 
The objective acoustic parameters in the first layer is $K_r(r_1)$, $K_\theta(r_1)$, $\rho(r_1)$. 
And the mean task is to find the right unit cell that has effective parameters identical to the objective ones. 
This can be realized by retrieving the bands of different unit cells and calculating their effective densities and modulus. 
\begin{figure}[ht]  
\begin{center}
\includegraphics[width=3in]{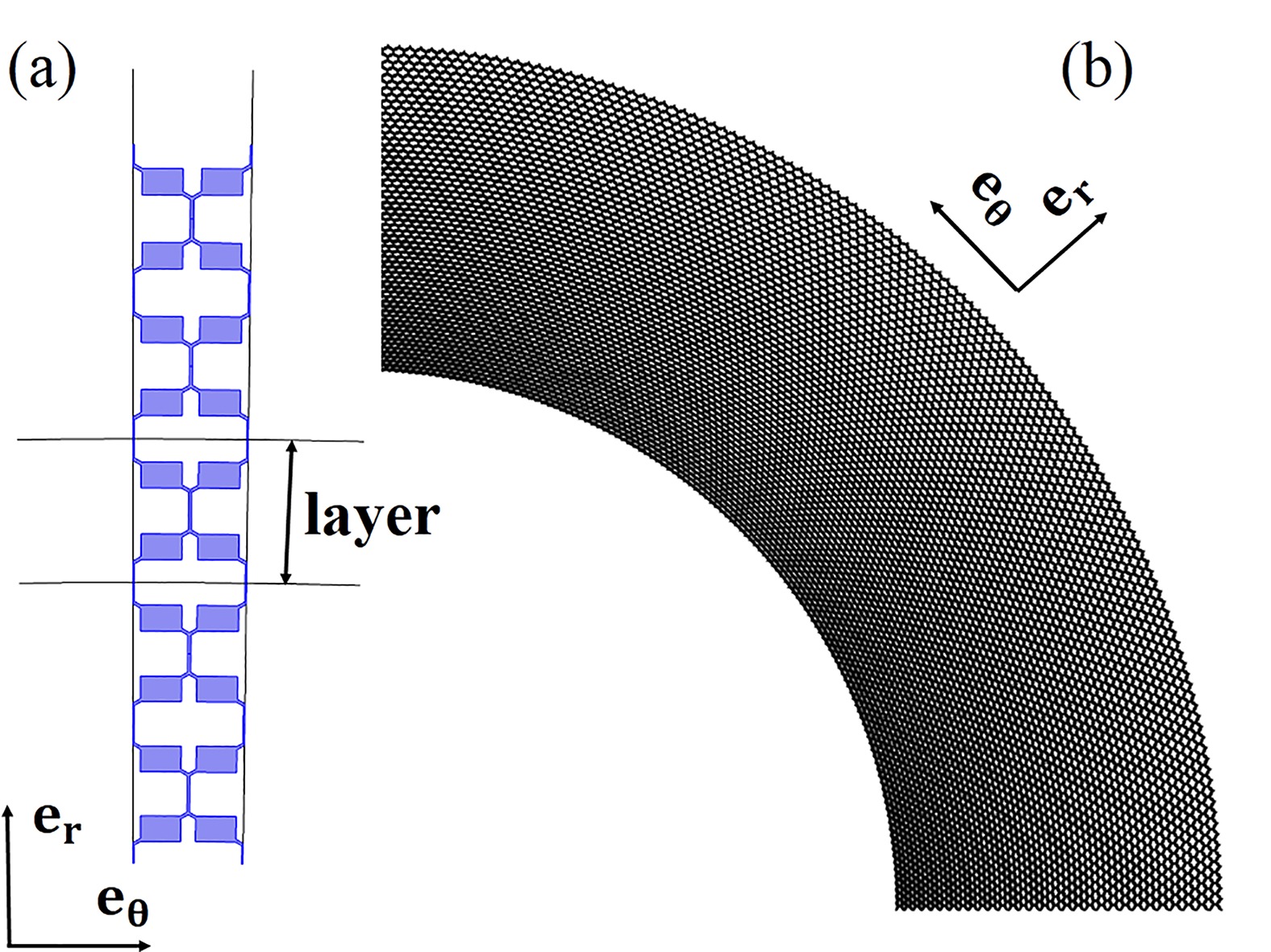}
\caption{\label{fig:FIG4}(Color online) (a) The microstructure of the first five layers in a sector. (b) Top view of the latticed $90^{\circ}$  PMB. }
\end{center}
\end{figure}
The structure of layers with microstructure is shown as in Fig.\ref{fig:FIG4}a. 
The ending of the first layer is the starting of the second layer.
The last layer ends at $r_2=1201.0$ mm.
The parameters of the unit cells in a sector is shown in Fig.\ref{fig:FIG3}b, while their corresponding effective acoustic parameters are shown as discrete symbols in Fig.\ref{fig:FIG3}a. 
It can be seen  that the effective parameters of the unit cells coincide with the required ones  well. 
We can also observe that the effective shear modulus of the structure is exceedingly small comparing with the compressive modulus. 
The bend structure is obtained by assembling the unit cells, shown as in Fig.\ref{fig:FIG4}b. 
The maximal period  is almost 15 mm which makes the unit cell at least 10 times smaller than the wavelength for frequencies under 8 kHz in water. 
Consequently, the layer can be regard as a homogeneous medium. 
And the effective acoustic parameters are expected to ensure the latticed PMB for smooth bending effect. 

\section{\label{simu}Simulation results}
\begin{figure*}[ht]
\begin{center}
\includegraphics[width=0.9\textwidth]{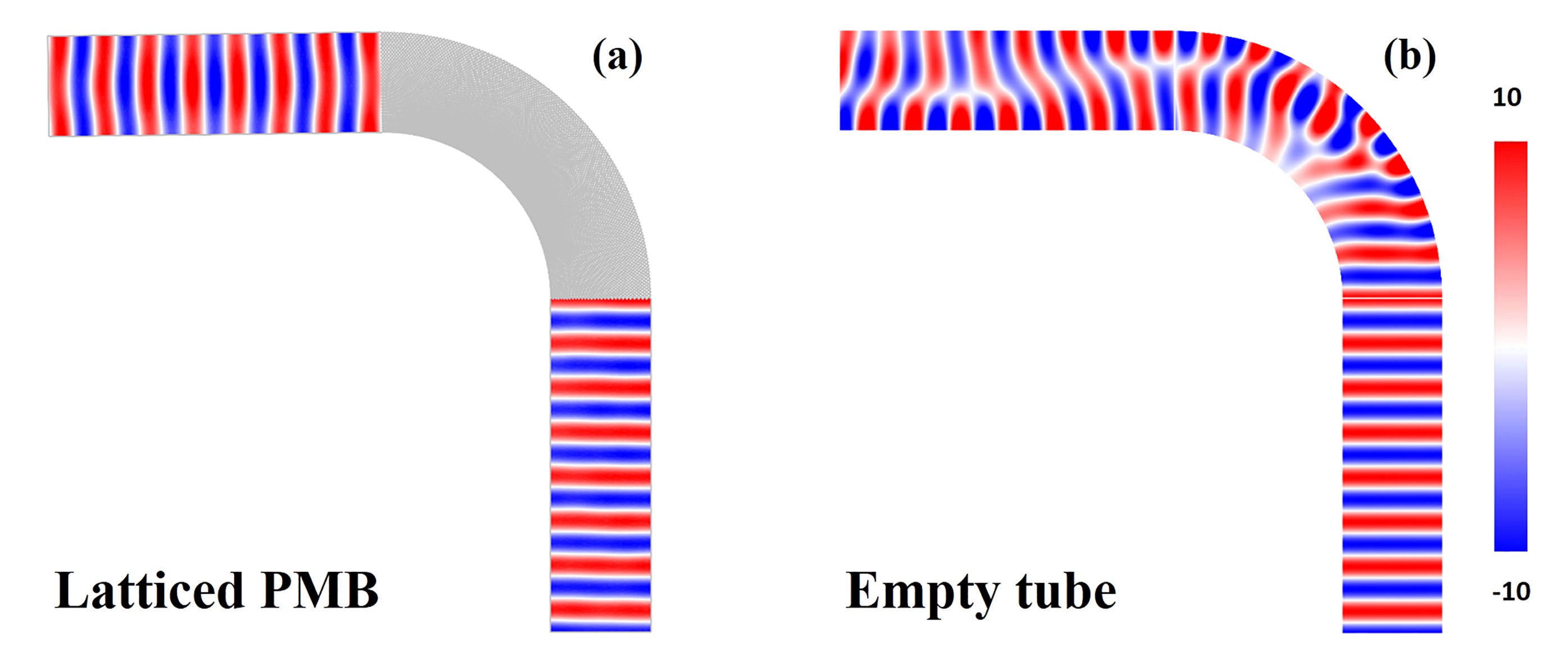}
\caption{\label{fig:FIG5} (Color online) Acoustic pressure field of the $90^\circ$ latticed PMB (a) and empty tube (b) for a plane wave incident from the bottom of the vertical waveguide at 7.5 kHz. }
\end{center}        
\end{figure*}  

The performance of the $90^\circ$ latticed PMB is simulated using finite element solver COMSOL Multiphysics. 
The incident and transmitted acoustic pressure fields of the latticed PMB are shown in Fig.\ref{fig:FIG5}a.  
A plane wave at 7.5 kHz is emitted from the bottom of the vertical waveguide. 
It shows that the wavefront in the horizontal waveguide almost keeps the same shape with that in the vertical waveguide. 
For comparison, an empty tube that has the same shape with latticed PMB is also simulated. 
The acoustic pressure fields are shown as in Fig.\ref{fig:FIG5}b. 
The wavefront in the horizontal waveguide in Fig.\ref{fig:FIG5}b is severely distorted.
Comparing Fig.\ref{fig:FIG5}a with Fig.\ref{fig:FIG5}b, it can be concluded that the latticed PMB can bend the wave smoothly at 7.5 kHz.
\begin{figure*}[ht]
\begin{center}  
\includegraphics[width=0.9\textwidth]{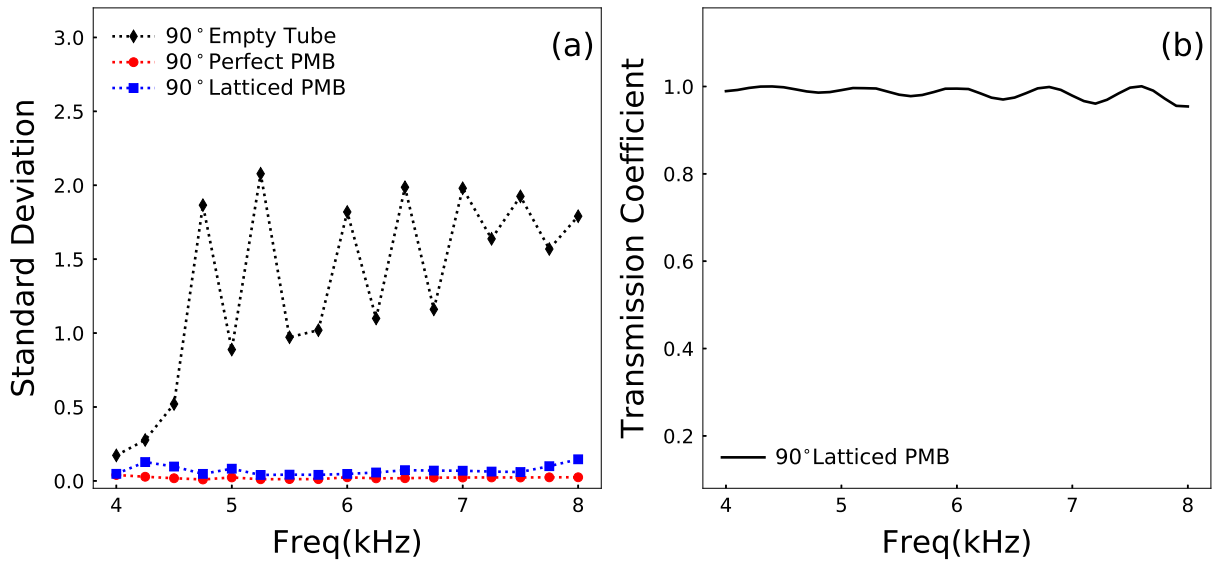}
\caption{\label{fig:FIG6} (Color online) (a) The standard deviations of a $90^\circ$ empty tube(black diamonds), a $90^\circ$ perfect PMB (red circles) and a $90^\circ$ latticed PMB. (b) The transmission of the $90^\circ$ latticed PMB. }
\end{center} 
\end{figure*}
   
    In order to evaluate the deformation of the transmitted wavefront quantitatively, we calculate the standard deviation(SD) of the phases  along the center line $L$(shown as dashed line in Fig.\ref{fig:FIG1}). 
The standard deviation of the phase $\sigma(\phi)$ is expressed as follows: 
\begin{equation}
\sigma(\phi)=\sqrt{\frac{\Sigma^{n}_{i}\left ( \phi_i-\overline\phi\right )^2}{n}}
\end{equation} 
where $\phi_i$ is the phase distribution along line $L$,  $\overline{\phi}$ is the average value of all the phases, and $n$ is the number of the total phase data. 
The value of the SD describes the degree  of that the wavefront is distorted, i.e., a large SD means a serious distortion.
Figure \ref{fig:FIG6}a shows the SDs of the  latticed PMB, empty tube and perfect PMB at the frequencies ranging from 4 kHz to 8 kHz.  
\begin{figure*}[ht]
\begin{center}
\includegraphics[width=0.98\textwidth]{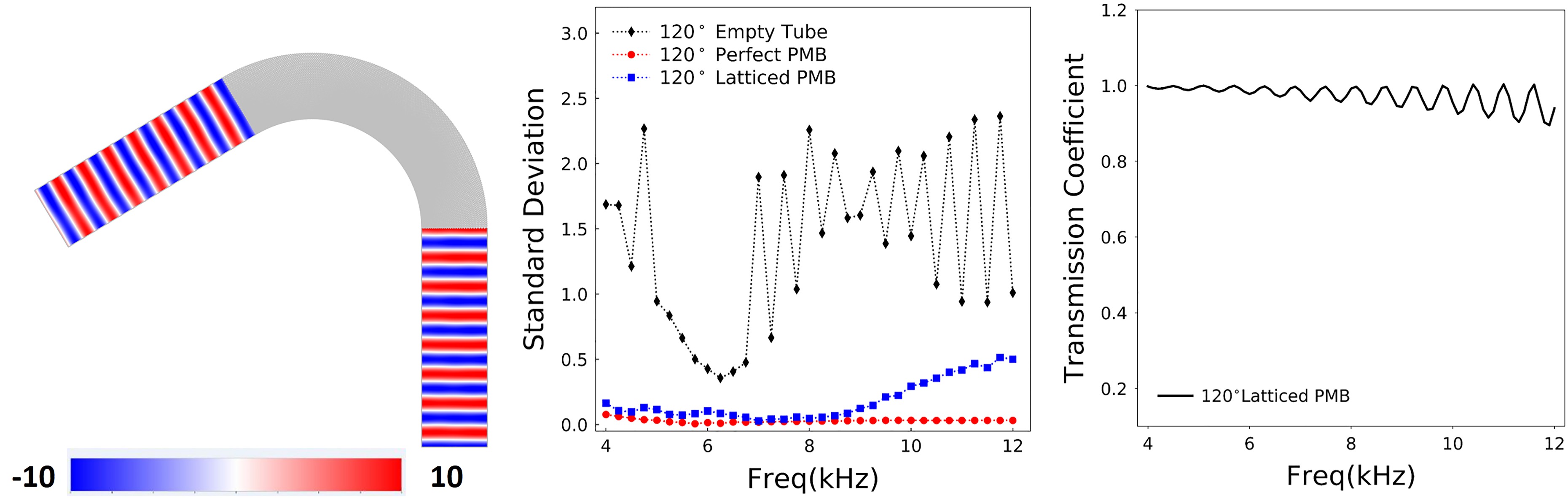}
\caption{\label{fig:FIG7}(Color online) (a) Acoustic pressure field of a  $120^\circ$ latticed PMB. (b) The standard deviations  for a $120^\circ$ empty tube(black diamonds), a $120^\circ$ perfect PMB (red circles) and a $120^\circ$ latticed PMB. (c) The transmission of the $120^\circ$ latticed PMB.}
\end{center}
\end{figure*}
It can be seen that the SDs of the perfect PMB are near to zero, which means the transmitted wavefront is almost not twisted. 
The SDs of the latticed PMB coincide with that of the perfect PMB very well.  
Thus, the latticed PMB is as effective as the perfect PMB at a broadband frequency domain. 

We also calculated the transmission coefficient of the PMB to evaluate  the degree of impedance matching. 
The transmission coefficient of $90^\circ$ latticed PMB is shown in Fig.\ref{fig:FIG6}b. 
It can be observed that the transmission coefficient is larger than 0.95. 
This implies that the acoustic impedance matches to water well. 
Thus, the $90^\circ$ latticed PMB has been verified to have a nice bending effect and be acoustic impedance matched to water.    
    
Since the latticed PMB is impedance matching in the azimuthal direction, the azimuthal length of the PMB will not influence the bending effect and transmission.   
Hence the latticed PMB can work at a wide bend angle range.
A $120^\circ$ latticed PMB is also designed and simulated.
The simulated results are shown in Fig.\ref{fig:FIG7}.  
The methods and evaluation for the $120^\circ$ latticed PMB is same with that of the $90^\circ$ one. 
It is obvious that the $120^\circ$ latticed PMB exhibits the bending effect as good as the $90^\circ$ latticed PMB.

\section{Conclusion}

In this article, we have designed and simulated a latticed pentamode material acoustic bend.
The required acoustic parameters that can keep the good bending effect and high transmission are obtained by a theoretical calculation.
The latticed  PMB is achieved by layered approximation with the 2D hexagonal PM unit cells that have the effective parameters according with the required ones. 
The unit cells are obtained by retrieving the bands and calculating their effective acoustic parameters.
The standard deviation of the transmitted phase is calculated to give a accurate analysis for the wave front, which shows the latticed PMB can keep the wavefront almost invariant.
And the high transmission confirms that the latticed PMB is impedance-matched to water well.
We hope this work can contribute to the researches on underwater communication and underwater detection.

\section*{Acknowledgments}

The authors sincerely acknowledge the financial support of the Youth  Innovation  Promotion  Association  CAS (Grant No. 2017029) and the National Natural Science Foundation of China (Grant  No.  11304351, 1177021304).
\bibliographystyle{plain}  
\bibliography{Metamaterials}
\end{document}